\documentclass[traditabstract]{aa} 
\usepackage{amssymb}
\usepackage{amsmath}
\usepackage{txfonts}
\usepackage{natbib}
\usepackage{graphicx}
\usepackage{lineno}
\bibpunct{(}{)}{;}{a}{}{,} 

\renewcommand{\phi}{\varphi}
\renewcommand{\theta}{\vartheta}

\newcommand{\referee}[1]{#1}

\begin{document}

\title{Anisotropic diffusion of galactic cosmic ray protons and their
  steady-state azimuthal distribution} \titlerunning{Anisotropic
  diffusion of galactic cosmic ray protons} \authorrunning{Effenberger et al.}

\author{F. Effenberger \and H. Fichtner \and
  K. Scherer \and I. B\"usching} \institute{Institut f\"{u}r
  Theoretische Physik IV, Ruhr Universit\"{a}t Bochum, 44780 Bochum,
  Germany, \email{fe@tp4.rub.de}} \date{Received ; Accepted}

\abstract{Galactic transport models for cosmic rays involve the
  diffusive motion of these particles in the interstellar medium. Due
  to the large-scale structured galactic magnetic field this diffusion
  is anisotropic with respect to the local field direction. We
  included this transport effect along with continuous loss processes
  in a quantitative model of galactic propagation for cosmic ray
  protons which is based on stochastic differential equations. We
  calculated energy spectra at different positions along the Sun's
  galactic orbit and compared them to the isotropic diffusion
  case. The results show that a larger amplitude of variation as well
  as different spectral shapes are obtained in the introduced
  anisotropic diffusion scenario and emphasize the need for accurate
  galactic magnetic field models.}


\maketitle

\section{Introduction}
Modelling the cosmic ray (CR) transport in the galaxy is a fundamental
topic in high energy astrophysics. Such studies are of great
importance in the analysis of CR origin like supernovae or pulsar
winds. The characteristics of the models depend on the properties of
the interstellar medium (ISM) through which the particles travel and
can thus give some insight into its fundamental constitution. For
example, the galactic magnetic field and its turbulent component are
connected to the transport parameters in such models. A reference on
general CR properties can be found, e.g., in the textbooks by
\citet{Berezinskii-etal-1990}, \citet{Gaisser-1990},
\citet{Schlickeiser-2002} or \citet{Stanev-2004}.

The basic propagation process of CR in the ISM is the diffusive motion
of the particles due to scattering at magnetic field
fluctuations. From numerous studies in heliospheric physics, it is
well known that the diffusive transport of energetic particles cannot
be described by a scalar diffusion coefficient but requires a
diffusion tensor which takes into account that parallel and
perpendicular diffusion are different
\citep[e.g.][]{Jokipii-1966,Potgieter-2011}. In galactic propagation
studies, however, anisotropic diffusion of CRs has been investigated
only for basic magnetic field configurations of partly localized
applicability, see, e.g., \citet{Chuvilgin-Ptuskin-1993},
\citet{Breitschwerdt-etal-2002}, \citet{Snodin-etal-2006} and
references therein.  While the latter authors were interested in the
consequences of anisotropic diffusion for energy equipartition,
\citet{Hanasz-Lesch-2003} and \citet{Ryu-etal-2003} analyzed its
importance for the Parker instability.

More recently, \citet{Hanasz-etal-2009} found that anisotropic
diffusion is an essential requirement for the CR driven magnetic
dynamo action in galaxies. \referee{Their assessment of the diffusion
  anisotropy is based on the full-orbit analysis performed by
  \citet{Giacalone-Jokipii-1999} who found the perpendicular diffusion
  to be significantly lower than the parallel one in a broad energy
  range and for both, isotropic and composite turbulence.} Moreover, a
recent derivation of the perpendicular diffusion coefficient for
galactic propagation, using the enhanced nonlinear guiding center
theory and a Goldreich-Sridhar turbulence model was performed by
\citet{Shalchi-etal-2010} and resulted in ratios between the parallel
and perpendicular diffusion coefficient which where much lower than
unity as well, namely $\kappa_{\perp}/\kappa_{\parallel}\approx
10^{-4}-10^{-1}$, depending on particle rigidity. \referee{An analysis
  for different turbulence spectra in the context of supernova remnant
  shock acceleration of CRs \citep{Marcowith-etal-2006}
  or their transport in chaotic magnetic fields
  \citep{Casse-etal-2002} yielded similar values for this ratio.}

Many popular models for galactic CR transport, however, include only a
single diffusion coefficient, like the GALPROP code
\citep{Strong-etal-2010}. Although \citet{Strong-etal-2007}
principally acknowledge that anisotropic diffusion is of importance,
they argue that due to large-scale fluctuations in the magnetic field
on scales of the order of 100pc, the global diffusion will be
spatially isotropic. Observations of the galactic magnetic field
indicate though that the field has a large-scale ordering with an
regular field strength with about the same magnitude as the turbulent
component \citep[see, e.g.,][]{Ferriere-2001}. A similar indication is
given by observations of external spiral galaxies
\citep{Beck-2011,Fletcher-etal-2011}, which show a global magnetic
field structure aligned to the spiral arm pattern.  Therefore, it must
be concluded that anisotropic diffusion can have an important effect in
galactic CR propagation.

Besides from its fundamental astrophysical relevance, the spatial
distribution of CR flux in the galaxy is also of interest in the
context of long-term climatology. \citet{Shaviv-2002} proposed a
CR-climate connection on the timescale of $10^8$ years due to the
transit of the solar system through the galactic spiral arms during
its orbit around the galactic center. The argument assumes that the
low-altitude cloud coverage increases due to an increased formation of
cloud-condensation nuclei when the CR flux is high
\citep{Svensmark-etal-2007}. Thus, an anti-correlation between
temperature and CR flux is to be expected and indeed reported by
\citet{Shaviv-Veizer-2003}. Most recently, \citet{Svensmark-2012}
found further evidence of a connection of nearby supernovae and their
CR output and life on Earth. More details on the CR-climate connection
can be found in \citet{Scherer-etal-2006}. Although we do not state
here that we adhere to this view in all aspects \citep[see, e.g., the
critical remarks in][]{Overholt-etal-2009}, we think that it gives
a further interesting motivation to study the galactic CR
distribution, and especially its longitudinal structure, in greater
detail.

The aim of this investigation is to calculate galactic CR spectra at
different positions along the Sun's orbit around the galactic center
and to analyse the influence of anisotropic diffusion on the
longitudinal cosmic ray distribution. We first present the underlying
propagation model and its relevant input, like the diffusion
tensor and its connection to the galactic magnetic field, the
three-dimensional source distribution of CR and its connection to the
spiral-arm structure and supernova (SN) occurrence, and loss-processes
in the ISM. We also introduce our numerical solution
method to the CR transport equation based on stochastic differential
equations. Finally, the calculated CR spectra and orbital flux
variations are discussed and conclusion are drawn. Some earlier
results on this topic can also be found in
\citet{Effenberger-etal-2011a}.

\section{The propagation model} 
The basic transport theory of CRs is described in many contemporary
monographs, e.g. \citet{Schlickeiser-2002}, \citet{Stanev-2004} and
\citep{Shalchi-2009}. Recently, \citet{Strong-etal-2007} have
surveyed the theory and experimental tests for the propagation of
cosmic rays in the Galaxy.  The considerations are
commonly based on the following parabolic transport equation
\citep[e.g.,][]{Ptuskin-etal-2006}:
\begin{eqnarray}
  \displaystyle
  \frac{\partial N}{\partial t} =
  \nabla\cdot\left(\hat{\kappa}\cdot\nabla\, N - \vec{u} N \right) 
  - \frac{\partial}{\partial p}\left[\dot{p} N
    - \frac{p}{3}(\nabla\cdot\vec{u}) N \right]
  + Q
\label{tp1}
\end{eqnarray}
where $N(\vec{r},p,t)=p^2f(\vec{r},p,t)$ is the differential intensity
of CRs and $f$ their phase space density, which is assumed to be
isotropic in momentum space.  As usual, $\vec{r}, p$, and $t$ denote
the location in space, momentum and time and we use a galactic
cylindrical coordinate system [$r,\phi,z$]. The source term $Q$
includes primary particle injection, which, in this study, is considered
to be only by supernovae and their remnant shock features. The spatial
diffusion, in general, should be described by a tensor, but in most
applications to galactic propagation so far, it is simplified to a
scalar coefficient $\kappa_s$, i.e. $\hat{\kappa} = (\kappa_{ij}) =
(\delta_{ij}\kappa_s)$ (see the discussion above). An ordered motion
of the ISM can be taken into account via the convection velocity
$\vec{u}$
\citep[e.g.,][]{Fichtner-etal-1991,Ptuskin-etal-1997,Voelk-2007}, but
is neglected for this study due to its decreasing importance for
higher CR energies. Continuous momentum losses are described by the
momentum change rate $\dot{p}$. Catastrophic loss processes like,
e.g., spallation do not apply, since in this study only galactic
protons are considered.

\subsection{The anisotropic diffusion tensor}
As discussed in the introduction, generally, the diffusion of CR in
magnetic fields with a prominent ordered field component is
anisotropic with respect to this field orientation, i.e. stronger in
field-parallel direction and weaker in the perpendicular
directions. This effect can be included in the propagation model by a
diffusion tensor which is locally, that is, in a field-aligned
coordinate system, diagonal:
\begin{equation}
\hat{\kappa}_{\rm{L}} =
\begin{pmatrix}
\kappa_{\perp 1} & 0 & 0 \\ 0 & \kappa_{\perp 2} & 0 \\ 0 & 0 & \kappa_{\parallel} \\
\end{pmatrix}
\end{equation}
Here, drift effects or aspects of non-axisymmetric turbulence
\citep{Weinhorst-etal-2008}, which could lead to off-diagonal elements
in the diffusion tensor, are neglected.

Since the CR transport is described in a global frame of reference
(i.e. the galactic frame with a cylindrical coordinate system in case
of this study) the field-aligned tensor has to be transformed to this
frame by the usual transformation
\begin{equation}
\hat{\kappa} = A \hat{\kappa}_{\rm{L}} A^{T} 
\end{equation}
This transformation is analogous to the Euler angle transformation
known from classical mechanics. The matrix $A = R_3 R_2 R_1$ describes
three consecutive rotations $R_i$ with $A^{-1} = A^T$ (since $A\in
\mathop{\mathrm{SO}}_3$). These rotations are defined by the relative
orientation of the local and the global coordinate system with respect
to each other.

If the two perpendicular diffusion coefficients are not equal, this
transformation is of particular importance in establishing the
appropriate orientation in the calculation of the global diffusion
tensor. Recently, \citet{Effenberger-etal-2012} established a
generalized scheme based on the local field geometry to account for
this. In the present study, however, both perpendicular diffusion coefficients
are set equal to reduce the set of unknown parameters (connected to
the unknown detailed turbulence properties in the ISM),
i.e. $\kappa_{\perp 1}=\kappa_{\perp 2}=\kappa_{\perp}$. Furthermore, since the
galactic magnetic field in consideration is to first order
parallel to the galactic disc (see the discussion in the following
subsection), the field tangential $\vec{e}_t$ and the z-axis
$\vec{e}_z$ unit vectors provide, together with the completing third
unit vector $\vec{e}_n = \vec{e}_z\times\vec{e}_t$, a well-defined
coordinate system. These unit vectors represent the columns of the
transformation matrix $A$.

In order to complete the model of the diffusive part of CR
propagation, the local tensor elements, i.e. $\kappa_{\parallel}$ and
$\kappa_{\perp}$ have to be defined. For the parallel diffusion
coefficient $\kappa_{\parallel}$, we assume the same broken power law
dependence as has been taken for the scalar diffusion coefficient in
\citet{Buesching-Potgieter-2008}, namely:
\begin{equation}
\kappa_{\parallel}=\kappa_0\left(\frac{p}{p_0}\right)^{\alpha} 
\end{equation}
with $\alpha=0.6 \, \textrm{for} \, p>p_0$, $\alpha=-0.48 \,
\textrm{for} \, p\le p_0$, $\kappa_0=0.027
\,\mathrm{kpc}^2/\mathrm{Myr}$ and $p_0 = 4\,$GeV/c. 

\referee{Originally, this particular break energy of 3-4 GeV was motivated to
fit the plain diffusion model results like in
\citet{Moskalenko-etal-2002} to the observed Boron to Carbon ratios. A
first refinement can be found in \citet{Ptuskin-etal-2006} and a more
rigorous study has been performed by
\citet{Shalchi-Buesching-2010}. They confirm, by inclusion of
turbulence dissipation effects and the replacement of the quasilinear
transport theory with a second-order diffusion theory, the possible
existence of such a turnover in the parallel diffusion coefficient.}

The perpendicular diffusion $\kappa_{\perp}$ is scaled to be a
fraction of $\kappa_{\parallel}$, i.e.
\begin{equation}
  \kappa_{\perp}=\epsilon\kappa_{\parallel}
\end{equation}
where the diffusion-anisotropy $\epsilon$ is assumed to be in the
range of 0.1 to 0.01 for galactic protons with GeV energies
\citep{Shalchi-etal-2010}. The actual variation of anisotropy with
energy is an interesting aspect, but for the relatively small energy
range up to 1~TeV, considered in this study, the anisotropy can be
regarded to first order as energy independent\referee{, thus following
  \citet{Giacalone-Jokipii-1999}. Since the latter author's finding of
  energy independence is valid for both isotropic \citep[see also the
  partially similar result by][]{Casse-etal-2002} and composite
  turbulence and given the to some extent different results found by
  \citet{Shalchi-etal-2010} assuming a Goldreich-Sridhar turbulence,
  conclusions on the detailed underlying turbulence properties should
  not be drawn on the basis of our study.}

\subsection{Galactic magnetic field models}
As soon as anisotropic diffusion is considered, the knowledge of the
large-scale magnetic field in the galaxy becomes important. Reviews on
this subject were written, e.g., by \citet{Beck-etal-1996},
\citet{Ferriere-2001} and \citet{Heiles-Haverkorn-2012}.  Pulsar
rotation measure data \citep{Han-etal-2006} give evidence for a
counterclockwise field orientation (viewed from the north Galactic
pole) in the spiral arms interior to the Sun's orbit and weaker
evidence for a counterclockwise field in the Perseus arm, see, however
the criticism by \citet{Wielebinski-2005}. In inter-arm regions,
including the solar neighbourhood, the data suggests that the field is
clockwise. \citet{Han-2006} proposed that the galactic magnetic field
in the disk has a bi-symmetric structure with reversals on the
boundaries of the spiral arms. Magnetic fields in the general class of
spiral galaxies were studied by, e.g., \citet{Wielebinski-Beck-2005}
and \citet{Dettmar-Soida-2006} and can be compared with that of our
own galaxy.

Fortunately, as long as drift effects are neglected, the actual
sign-dependent orientation is of no relevance for the diffusion along
and perpendicular to the magnetic field, which enables us to employ a
simplified model of the galactic magnetic field that has no field
reversals and is aligned to the spiral arm structure in the
disc. Neglecting furthermore its weak halo-component, a simple model
for the mean galactic magnetic field in cylindrical coordinates is
given by
\begin{equation}
\mathbf{B} = B_0 (\sin\psi\, \mathbf{e}_{r} + \cos\psi\, \mathbf{e}_{\phi}) \frac{1}{r} \exp\left(-\frac{z^{2}}{2\sigma_{z,m}^2}\right)
\end{equation}
which is divergence free by construction. Here, $\psi$ is the
counter-clock logarithmic spiral arm pitch-angle, which has an
approximate value of $\psi=12^\circ$, according to the meta-study by
\citet{Vallee-2005}. The same spiral arm parametrization is employed
for the source distribution function discussed in the following
section. Note, that the halo-scale parameter $\sigma_{z,m}$ is not
relevant in this context, since only the magnetic field direction is
used for the construction of the diffusion tensor.

\subsection{Source distribution}
\begin{figure}[!pt]
   \includegraphics[width=0.48\textwidth]{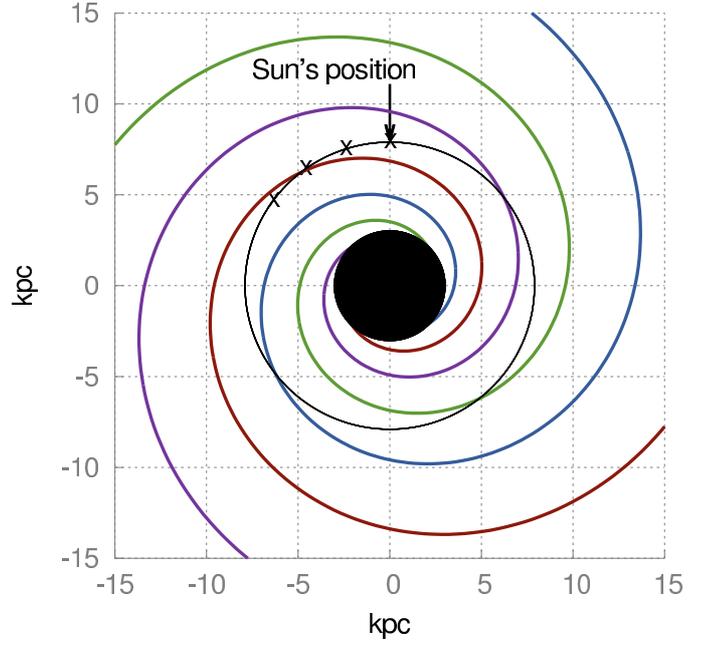}
   \caption{Orientation of the galactic spiral arms in the present
     model. Norma, Scutum, Saggitarius, and Perseus arms are colored
     by green, blue, red, and purple, respectively. The black line
     shows the solar orbit and the galactic center region is marked in
     black as well. The four x-markings indicate the positions at
     $90^{\circ}$, $108^{\circ}$, $126^{\circ}$, and $144^{\circ}$, where
     the CR spectra have been calculated (see Section~\ref{sec:results}).}
   \label{fig:orientation}
\end{figure}
\begin{figure}[!pt]
   \includegraphics[width=0.48\textwidth]{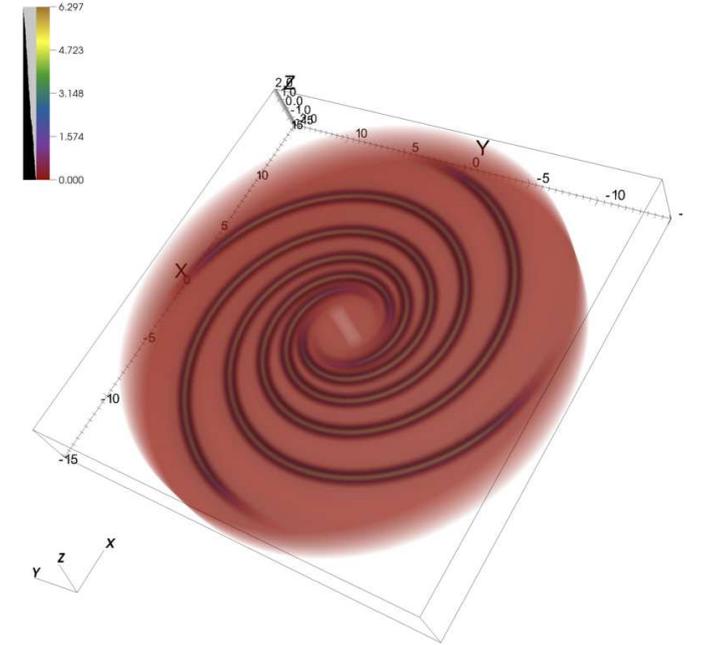}
   \caption{Volume rendering visualisation of the input source
     distribution of CRs. The coloring indicates the source strength
     from red (low) to yellow (high) in arbitrary units.}
   \label{fig:sources}
\end{figure}
For the injection of CRs we assume a source distribution which
follows the galactic spiral arm structure, where supposedly most of
the supernovae (SN) occur. As a basis we take the same spiral arm
model as mentioned above, i.e. the model established by
\citet{Vallee-2002,Vallee-2005}, which consists of four logarithmic
and symmetrically positioned arms. Around these, we take a Gaussian
shape (analogous to the approach in \citet{Shaviv-2003}) to yield an
analytic expression for the source term $Q$, by summing up over all
four arms ($n\in$ \{1,2,3,4\}):
\begin{equation}
q_n = Q_0\,p^{-s}\,\exp\left(-\frac{(r-r_n)^{2}}{2\sigma_r^2}
 - \frac{z^{2}}{2\sigma_z^2}\right)
\end{equation}
with $r_n=r_0 \exp({k(\phi + \phi_n)})$. $\phi_n$ introduces the
symmetric rotation of each arm by $90^{\circ}$, i.e. $\phi_n=
(n-1)\pi/2$. $k=\cos\psi$ with $\psi=12^{\circ}$ is the constant
pitch-angle cosine of the spiral arms. Figure~\ref{fig:orientation}
illustrates the orientation of the spiral arms relative to the Sun's
position and orbit. We take $\sigma_r = \sigma_z = 0.2$ kpc to have a
reasonable inter-arm separation, while $r_{0} = 2.52$ kpc according to
{Vall{\'e}e}'s model.
\begin{figure*}[!pt]
   \includegraphics[width=0.48\textwidth]{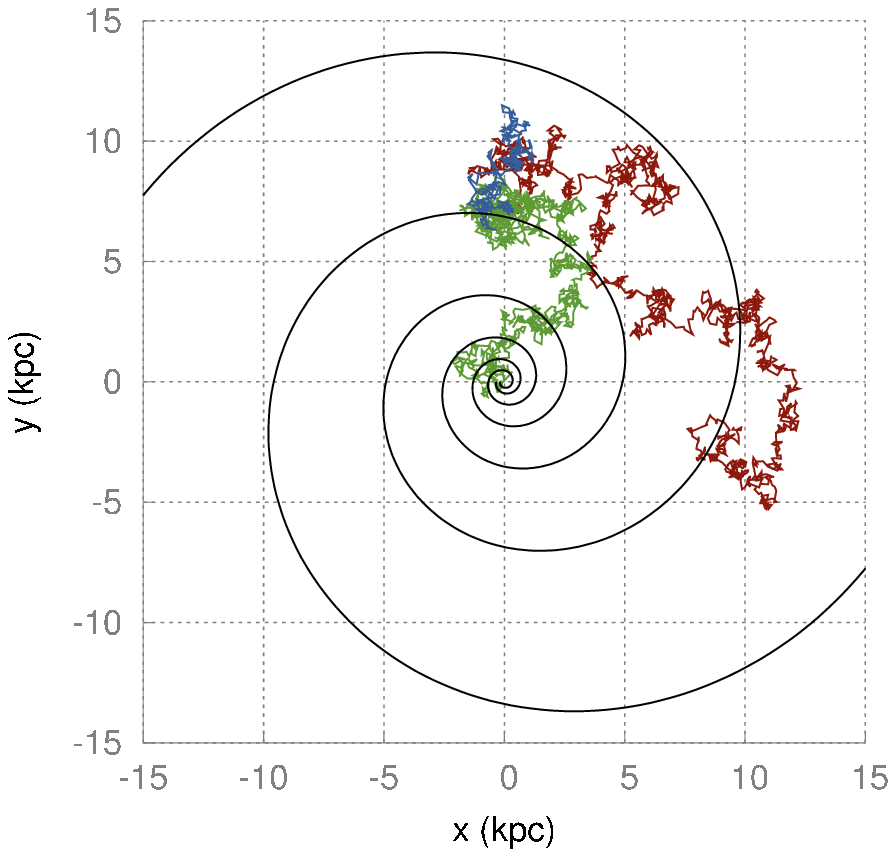}
   \includegraphics[width=0.48\textwidth]{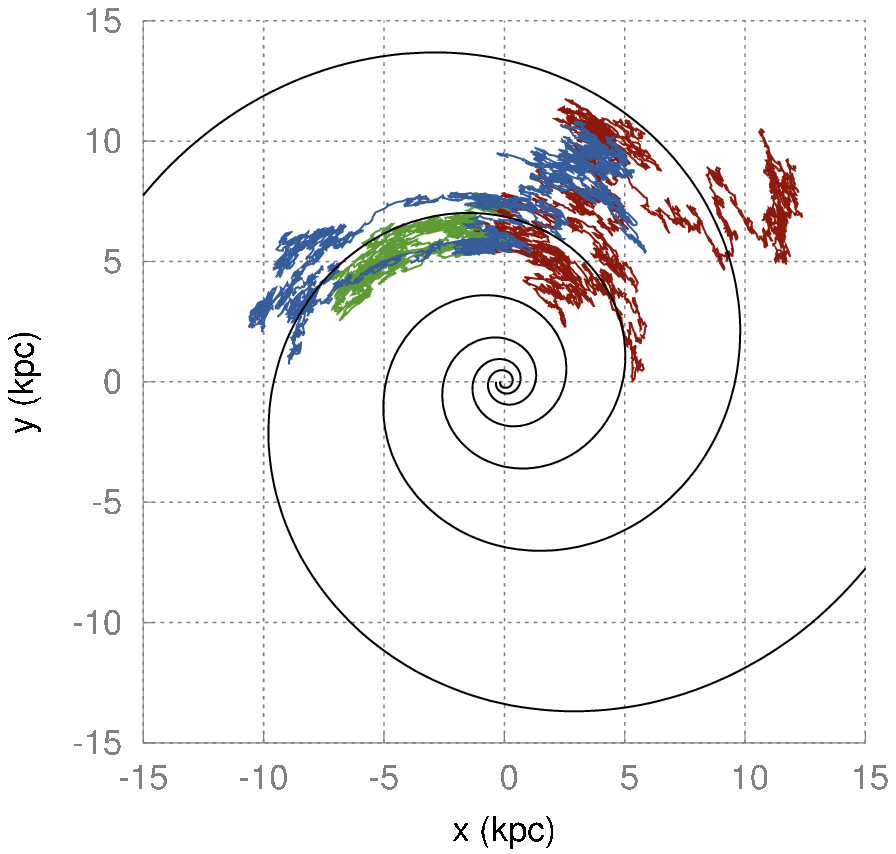}
   \caption{Sample paths of pseudo-particles in the galactic magnetic
     spiral field, coloured by three different colours for three
     particles, each starting at the same point in phase space
     (i.e. Earth's position at 1~GeV) and projected onto the galactic
     plane. The black lines show integrated magnetic field lines to
     illustrate the magnetic field orientation. The left panel shows
     sample paths for isotropic diffusion, with no visible effect of
     the magnetic field orientation. The right panel illustrates the
     preferential diffusion along the magnetic field for a simulation
     with anisotropy $\epsilon = 0.1$. Note that the exit point of the
     red particle in the right panel is actually the radial boundary,
     while the other particles all exit through the halo's z-boundary
     (not visible).}
   \label{fig:paths}
\end{figure*}
The model galaxy has the often assumed
cylindrical shape with a radius of 15~kpc and a height of 4 kpc
\citep{Buesching-etal-2005} and the Sun's orbit is at a radius of
$r_{\sun}=7.9$~kpc. The average spectral index $s$ of the sources' power law
injection in momentum is set to $s=2.3$ in agreement with recent
estimates on CR source spectra \citep[see
e.g.][]{Putze-etal-2011,Ave-etal-2009}.  Figure~\ref{fig:sources}
gives a visualisation of this source distribution. The overall source
strength $Q_0$ is a free parameter which can be fitted to a given
reference like a local interstellar spectrum (see also the discussion
in Section~\ref{sec:results}).

\subsection{Loss processes}
The two most dominant loss processes for CR protons during their
propagation through the ISM are energy losses due to pion-production
for relativistic energies and ionization processes in the ISM plasma for
lower energies \citep[see Fig.1 in][]{Mannheim-Schlickeiser-1994}.

According to Chapter~5 in \citet{Schlickeiser-2002} the pion losses
can be approximated for Lorentz factors $\gamma \gg 1$ as
\begin{equation}
-\left(\frac{\mathrm{d}\gamma}{\mathrm{d}t}\right) = 1.4\cdot 10^{-16}(n_{HI} +
2n_{H_2})\,A^{-0.47}\gamma^{1.28}\mathrm{s}^{-1}
\end{equation}
where we assume a z-dependent ISM gas density with $(n_{HI} +
2n_{H_2}) = \frac{1.24}{\cosh(30\,z)}$ \citep[in units of particles
per cm$^3$ and $z$ in kpc,][]{Buesching-Potgieter-2008}. The mass
number $A$ of a proton is just unity. A similar formula for the
ionization losses is given by
\begin{equation}
-\left(\frac{\mathrm{d}p}{\mathrm{d}t}\right) = 3.1 \cdot
10^{-7}\,Z^2\,n_{e}\frac{\beta}{x_m^3 + \beta^3}\mathrm{eV}\,\mathrm{c}^{-1}\,\mathrm{s}^{-1}
\end{equation}
where the electron density is the same as the above gas density and
the charge state of protons $Z$ is equal to $1$. For the purpose of
this study, the velocity factor $\beta = v/c$ of the particles is
always much greater than the thermal electron $\beta_{e}$ which is
related to $x_m$ by $x_m = (3/4\pi^{1/2})^{1/3} \beta_{e}=
1.10\,\beta_{e}$. This means that the momentum loss rate scales
approximately as $p^{-2}$. The total momentum loss rate entering
Equation~\ref{tp1} is the sum of both loss processes.

\subsection{The numerical solution method based on stochastic differential equations}
To solve the transport equation for the problem setup introduced in
this study, we employ a numerical solution scheme based on the
It$\bar{\text{o}}$ equivalence of a Fokker-Planck type equation and
corresponding stochastic differential equations (SDEs) involving a
random Wiener process. This method has become increasingly popular in
CR transport studies because of its numerical simplicity and
conformance with modern computer architecture, i.e. its
straightforward parallelization and scalability. Mentioning only a few
examples, a starting point for heliospheric studies of this kind can
be found in the paper by \citet{Zhang-1999a} where he applied the
method to CR modulation. More recently, \citet{Pei-etal-2010} and
\citet{Strauss-etal-2011} applied SDEs in a more comprehensive
heliospheric model. \citet{Farahat-etal-2008} applied SDEs for a CR
propagation study in the Galaxy and, e.g., \citet{Marcowith-Kirk-1999}
as well as \citet{Achterberg-Schure-2011} calculated the shock
acceleration of energetic particles.

\begin{figure}[!pt]
   \includegraphics[width=0.47\textwidth]{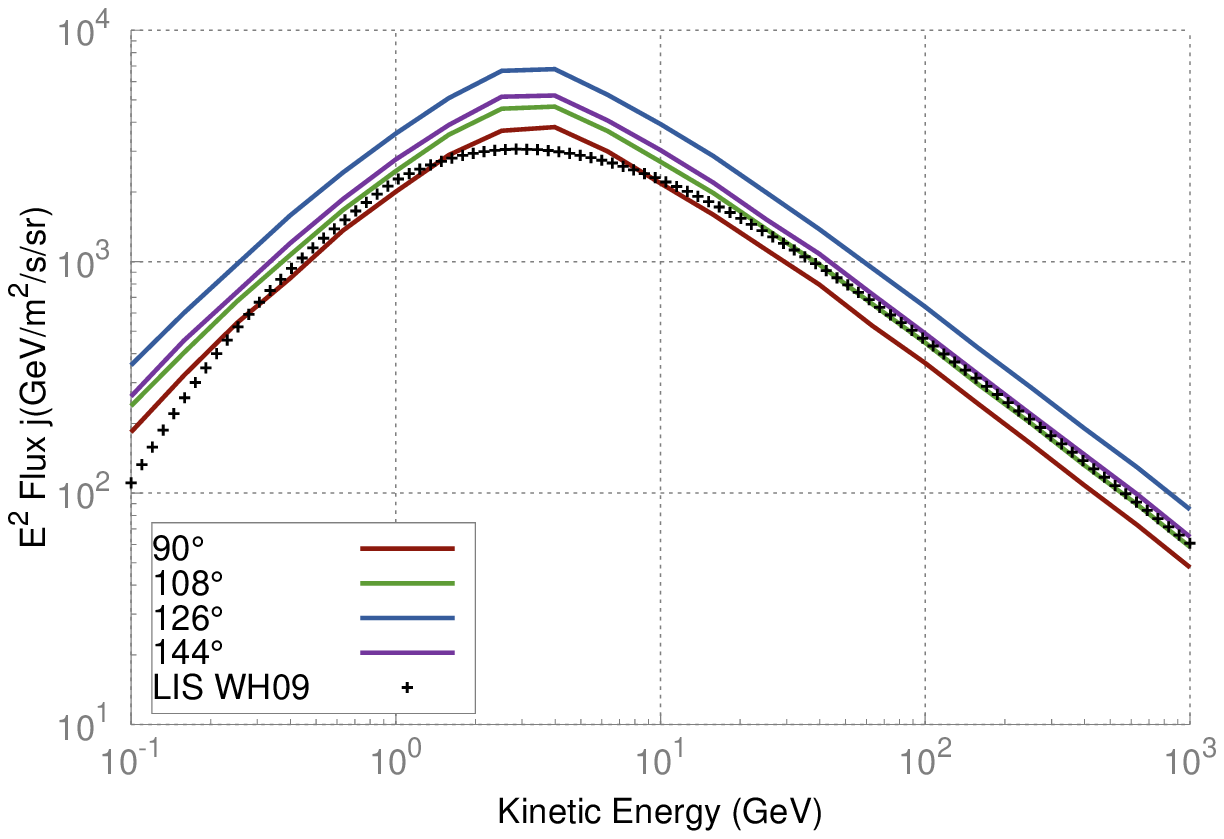}\\
   \includegraphics[width=0.47\textwidth]{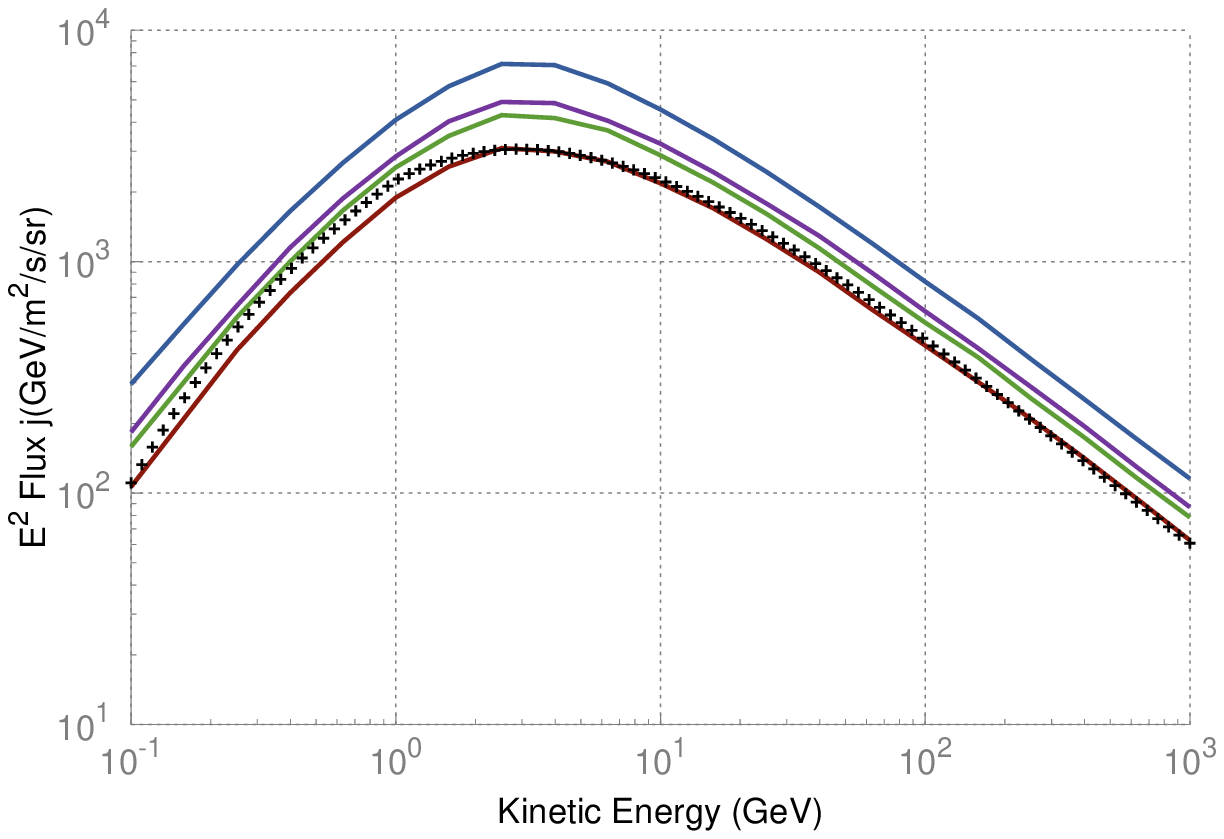}\\
   \includegraphics[width=0.47\textwidth]{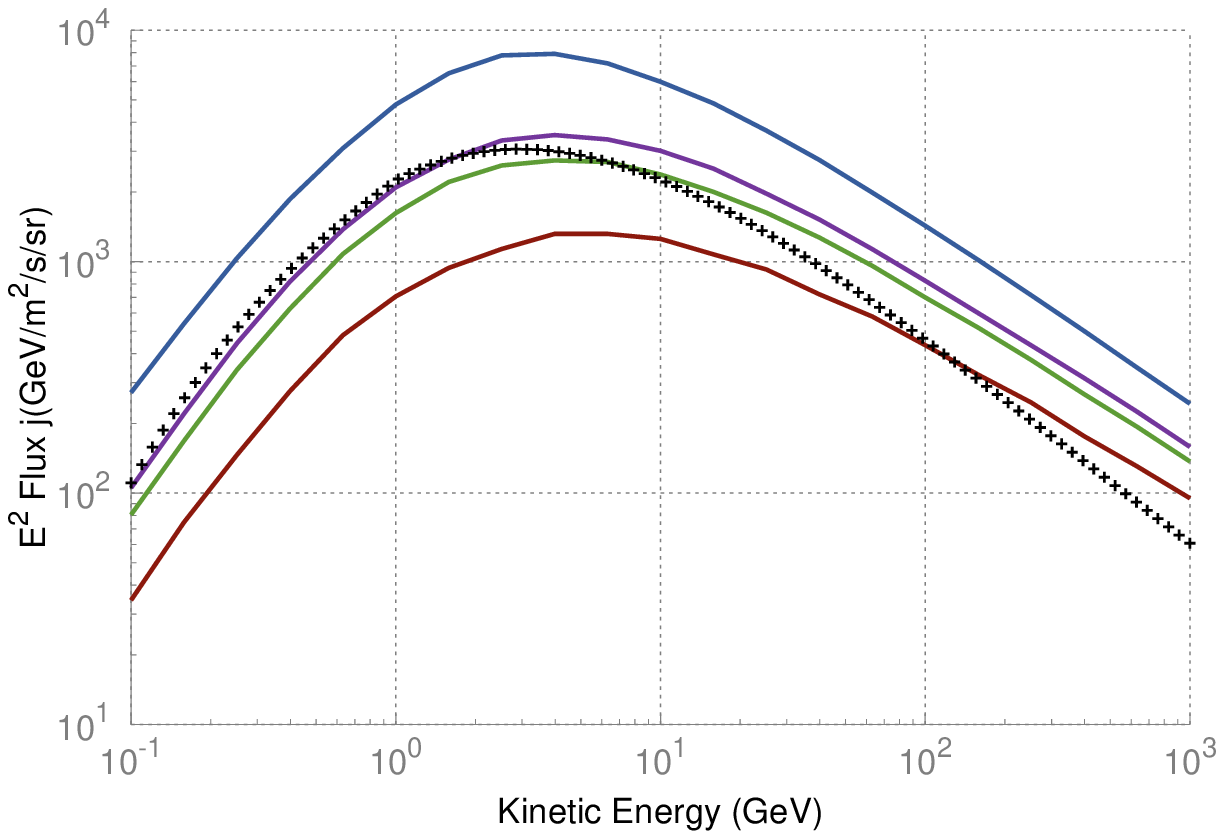}
   \caption{Calculated CR proton spectra (multiplied by $E^2$) at four
     different positions along the Sun's orbit (see
     Fig.~\ref{fig:orientation} for the respective locations) in the
     galactic plane ($z=0$), where $90^{\circ}$ corresponds
     approximately to the current solar system position and
     $126^{\circ}$ is inside of the Saggitarius arm. The upper panel
     shows the spectra for isotropic diffusion, while the middle and
     lower panel include anisotropic diffusion with $\epsilon = 0.1$
     and $\epsilon = 0.01$, respectively. The latter spectra have been
     rescaled by factors of $0.15$ and $0.05$ to account for the
     higher overall flux due to the stronger confinement in the disk
     in contrast to the isotropic case. The LIS from
     \citet{Webber-Higbie-2009} is plotted for comparison (black
     crosses).}
   \label{fig:spectra}
\end{figure}

The basic idea in SDE schemes is to trace pseudo-particle trajectories
from their origin forward in time or, alternatively, integrate
backwards in time from the phase space point of interest. The particle
trajectory is given by the integral of an SDE of the following form
\begin{equation}
dx_i = A_i(x_i)ds + \sum_jB_{ij}(x_i)dW_j
\end{equation}
where the relation $\hat{B}\hat{B}^T = 0.5\,\hat{\kappa}$ has to be
fulfilled, that is, a root for the diffusion tensor $\hat{\kappa}$ has to be
determined. Here, $dW_j$ is a (multidimensional) Wiener process
increment, which has a time-stationary normal-distributed probability
density with expectation value 0 and variance 1. The deterministic
part is directly related to the convection velocity in the transport
equation, i.e. $A_i = -u_i$. Numerically, this SDE is integrated via a
simple Euler-forward scheme and the Wiener-process is simulated with
the Box-Muller method \citep[e.g.,][]{Box-Muller-1958} by using
\begin{equation}
dW_i(s) = \eta(s)\sqrt{ds}
\end{equation}
where $\eta(s)$ is equivalent to a Gaussian distribution
$\mathcal{N}(0,1)$. The necessary random numbers are generated with
the MTI19937 version of the so called Mersenne Twister
\citep{Matsumoto-Nishimura-1998}. The integration parameter $s$ is
related to physical time by
\begin{equation}
t = t_0 - s
\end{equation}
where $t_0$ is the final time for the backward method. The source
contribution to the individual particle trajectory is added up by a
path amplitude. Finally, in case of the backward method, all particle
trajectories are weighted together to yield the resulting phase space
density (i.e. the solution to the associated Fokker-Planck equation)
at the starting phase space point. The boundary and initial conditions
can be accounted for in the weighting, but for this study they are
simply zero (corresponding to an escaping boundary condition for the
CRs). We only apply the backward method in this study, since it is
well-suited for the given problem. For more details on the numerical
scheme and especially on the determination of the root of the
diffusion tensor, we refer the reader to \citet{Kopp-etal-2012} and
\citet{Strauss-etal-2011} where the basis of the code used in this
study is discussed in greater detail.

Exemplary pseudo-particle trajectories are shown in
Fig.~\ref{fig:paths}. There, the additional information contained in
SDE calculations becomes obvious. The pseudo-particles' paths follow the
field lines during their stochastic motion as soon as anisotropic
diffusion becomes relevant. Consequently, the modification to the
diffusion process becomes directly visible in such
trajectories. However, one has to keep in mind that these are not real
particle trajectories or gyro-center motions, but only tracers of the
phase space of the diffusion-convection problem.

\section{The resulting spatial and spectral CR distribution}
\label{sec:results}
We calculated CR proton spectra within the introduced model at four
different positions along the Sun's galactic orbit. The positions are
indicated in Figure~\ref{fig:orientation}. We took a very long
integration time ($t_{0}\approx 10000$~Myrs) to assure that we
approached a steady state situation, which is confirmed by checking
that all particles have exited the computational domain. For each
phase-space point, $10^4$ pseudo particle trajectories have been
computed. A comparison between the calculated spectra in the case of
pure isotropic diffusion (upper panel) and two anisotropic case, with
weak ($\epsilon = 0.1$, middle panel) and strong ($\epsilon = 0.01$,
lower panel) diffusion anisotropy, is shown in
Figure~\ref{fig:spectra}. For comparison, the local interstellar
spectrum (LIS) given by \citet{Webber-Higbie-2009} is included in the
plots (WH09 hereafter). We have used the parametrization given in
\citet{Herbst-etal-2010}, where a comparison between a few proposed
LIS can be found as well. In face of the still imprecisely known
modulation effects on measured spectra inside the heliosphere
\citep[see, e.g.,][]{Florinski-etal-2011,Scherer-etal-2011}, such an LIS parametrization
can give only a rough orientation on what to expect for galactic CR
propagation studies. Our results have been rescaled to fit
approximately to the WH09 LIS in the isotropic diffusion case, by
accounting for the free parameter $Q_0$ in the source strength. The
anisotropic spectra have been rescaled again, respectively. To yield
the good agreement shown in both upper panels of
Figure~\ref{fig:spectra} between the calculated spectra and the WH09
LIS, the break in the diffusion coefficient introduced above as well
as both continuous loss processes are required. The inclusion of the
latter is an improvement over earlier studies, like
\citet{Buesching-Potgieter-2008}, where only a parametrized
catastrophic loss term was considered.

The spectra for different positions along the Sun's galactic orbit
show only very little variation in the isotropic diffusion
case. Particularly, the variation is largely independent of energy
over the entire energy range considered.  In contrast to this, the
variation is much stronger for the anisotropic case, depending on the
imposed diffusion anisotropy. The differences are, in these
cases, dependent on energy as well. For high energies, the spectra
start to converge again towards the isotropic differences. This is due
to the increasing dominance of escape losses for these high
energies. For lower energies, the pion and ionization losses are much
more important than in the isotropic case, because the confinement
time of CRs is longer as a result of the reduced diffusion
perpendicular to the disk.

\begin{figure*}[!pt]
   \includegraphics[width=0.48\textwidth]{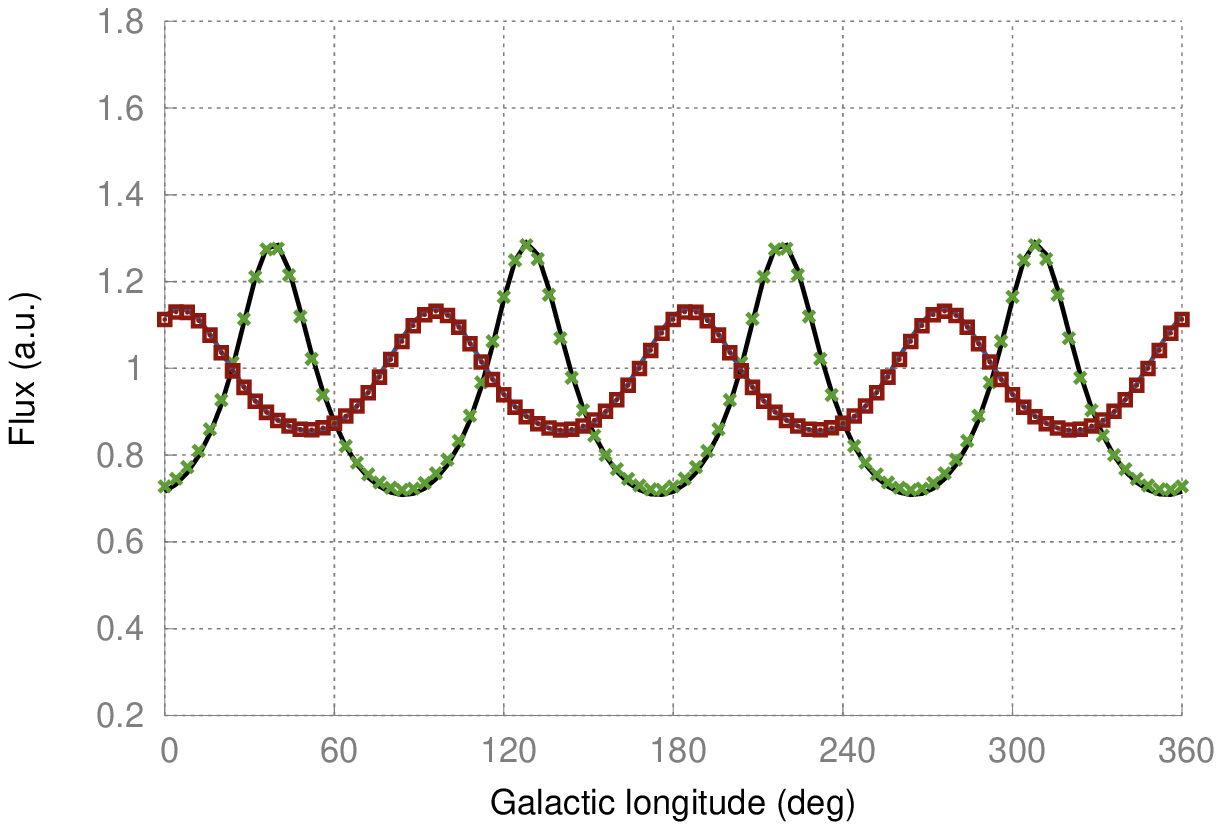}
   \includegraphics[width=0.48\textwidth]{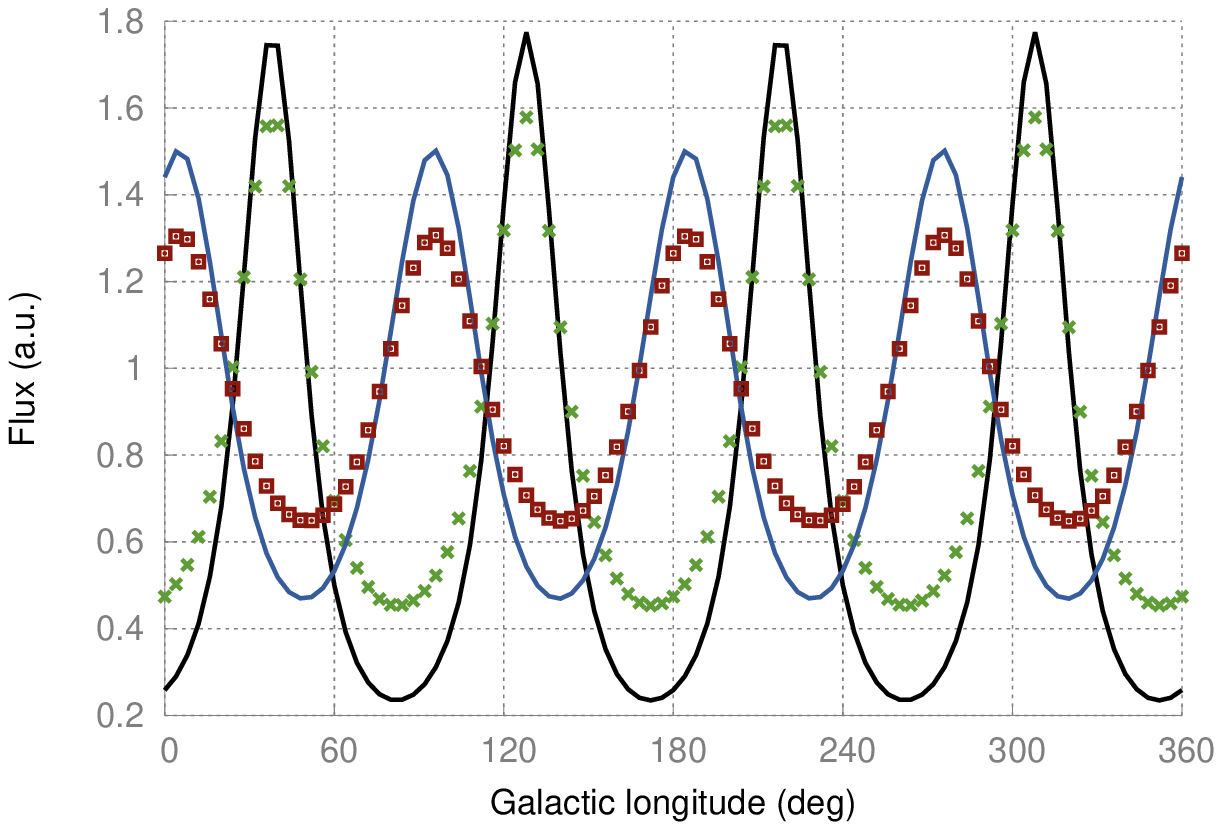}
   \caption{Orbital variation of CR proton flux along the Sun's orbit,
     plotted against longitude (with Sun's position at
     $90^{\circ}$). All curves are rescaled to an offset value of 1
     for comparison. The black line and green crosses show the
     variation for an orbit at $r=7.9$\,kpc and $z=0$ for 1\,GeV and
     100\,GeV, respectively. The blue line and red boxes show the same
     energies for an orbit at $r=5.$\,kpc. The left panel shows again
     the case of isotropic diffusion, while the right panel has $\epsilon
     = 0.01$ and the variation is much more pronounced.}
   \label{fig:orbits}
\end{figure*}

Notably, the spectrum at Earth for the weak anisotropic case fits even
better to the reference LIS than the pure isotropic result, which shows
that, depending on the overall parameter set, anisotropic diffusion
scenarios can improve on the model results of conventional studies.
In this context, one has to keep in mind however, that the precise
form of the low energy LIS and its connection to the galactic spectrum
on a kpc scale is yet unclear and depends on modulation effects in the
heliosphere as well as similar effects in the local solar system
environment \citep[see, e.g., the discussion
in][]{Scherer-etal-2011}. Furthermore, the assumed break in the
diffusion coefficient may be different or even absent in a more
complete propagation scenario, since up till now it is mainly
phenomenologically motivated, to yield the expected local spectra.

The spectra for the strong anisotropic case ($\epsilon=0.01$) show
some significant deviations from the expected spectral shape due to
the largely increased relative importance of the loss processes,
resulting e.g. in a flatter high energy spectrum. In the context of
the model setup of this study, this means that such a high diffusion
anisotropy is probably unrealistic. Nevertheless, we included this
case since it shows the resulting large orbital variation at lower
energies (see also Fig.~\ref{fig:orbits}) where the spectral shape is
still unclear. In addition, models with different structures in the
galactic halo, namely with different gas densities and halo heights,
as well as a possible magnetic field component perpendicular to the
disc, may alter the resultant spectra further, due to a changed
influence of the loss processes. These aspects could be further
clarified in a subsequent study which takes different CR species and
more sophisticated magnetic field models into account.

In Figure~\ref{fig:orbits} the orbital flux variation along the Sun's
orbit is plotted against longitude, to illustrate further the variation
with longitudinal position. We consider two different energies, namely
1~GeV and 100~GeV and two galactic distances of $r=5$~kpc and
$r_{\sun}=7.9$~kpc, again for isotropic (left panel) and strong
anisotropic ($\epsilon = 0.01$, right panel) diffusion. The inclusion
of a second radius at only 5~kpc is motivated by the recent claim that
the Sun may have migrated outwards during its lifetime in the galaxy
\citep{Nieva-Przybilla-2011}. It can be seen that the variation is
much weaker in a closer galactic orbit and has a different phase, as a
result of the smaller inter-arm separation. For all cases, the overall
shape of the variation is not a simple sinusoidal profile due to the
non-perpendicular transit of the Sun through the arms (see again
Figure~\ref{fig:orientation}). The amplitude of variation is much more
pronounced for the anisotropic case, that is, it can be as large as a
factor of $6$, while in the isotropic case it is only a factor of
about $2$. An increased diffusion anisotropy $\epsilon$ will increase
this difference even further.

\section{Conclusions}
In this study the effects of anisotropic diffusion of galactic CR
protons have been analyzed. For the solution of the steady-state
diffusion equation a numerical method based on stochastic differential
equations has been used, which also accounts for energy loss
processes. The computed spectra along the Sun's galactic orbit show
larger variations for the anisotropic cases when compared to the
scalar diffusion model. Furthermore, for the chosen parameters, a
moderate diffusion anisotropy ($\epsilon =
\kappa_{\perp}/\kappa_{\parallel}= 0.1$) leads to a result which, in
our setup, is more consistent with recent estimates of the local
interstellar proton spectrum than the results for purely isotropic
diffusion.

We therefore conclude that the diffusion tensor as well as
the CR source distribution is an important feature in determining the
solution of the transport equation in a three-dimensional model of
galactic CR propagation. This result fits well into the findings by
\citet{Hanasz-etal-2009} claiming that anisotropic diffusion is an
essential requirement for the CR driven galactic dynamo effect.
Additionally, these results imply that in the context of the
recently proposed CR-climate connection \citep{Shaviv-Veizer-2003} the
expected CR flux variation may be even larger than previously
estimated, although at present, this topic is still highly
speculative.

We emphasize that the introduced model indicates further research
opportunities by including additional effects like a variable
spectral source index and time-variable CR sources, depending on
supernova type. The necessary time-dependent calculations are in
principle possible with the current model setup, and the earlier work
by \citet{Buesching-Potgieter-2008} shows some results on this
already. Furthermore, the consideration of additional CR species and
their relevant loss processes may yield further insight to constrain
the transport parameters. Finally, this study points at the necessity
to acquire more detailed models of the galactic magnetic field, to
asses its impact on the transport processes of CRs.

\begin{acknowledgements}
  The work was carried out within the framework of the `Galactocauses'
  project (FI 706/9-1) funded by the Deutsche Forschungsgemeinschaft
  (DFG) and benefited from the DFG-Forschergruppe 1048 (project FI
  706/8-1/2), the 'Heliocauses' DFG-project (FI 706/6-3) and funding
  by the German Ministry for Education and Research (BMBF) through
  ``Verbundforschung Astroteilchenphysik'' grant 05A11PC1. We thank
  A. Kopp, M. Pohl and R. Schlickeiser for useful discussions with
  regard to the modelling approach in this study and an anonymous
  referee for helpful comments.
\end{acknowledgements}


\end{document}